\documentclass[11pt,a4paper]{article}
\usepackage{amsfonts,amssymb}
\usepackage{cite}
\usepackage[cp1251]{inputenc}
\usepackage[english]{babel}
\usepackage{amsmath}%
\usepackage[unicode]{hyperref}

\newtheorem{Theorem}{Theorem}
\newtheorem{Lemma}{Lemma}
\newtheorem{Spisok}{List}
\def\spisok#1{\begin{gather}#1\end{gather}}
\def\eq#1{\begin{equation}#1\end{equation}}
\def\eqs#1{\begin{equation}\begin{split}#1\end{split}\end{equation}}
\def\seqs#1{\begin{equation*}\begin{split}#1\end{split}\end{equation*}}
\def\seq#1{\begin{equation*}#1\end{equation*}}
\def\qed{\vrule height0.6em width0.3em depth0pt}
\font\Sets=msbm10
\def\integer {\hbox{\Sets Z}}

\title{\bf Generalized symmetry classification of discrete equations of a class depending on twelve parameters}
\author{{\bf R.N. Garifullin}\\{\sl E-mail: \url{rustem@matem.anrb.ru}}\\{\sl URL: \url{http://matem.anrb.ru/garifullinrn}}
\\ Ufa Institute of Mathematics, Russian Academy 
of Sciences,\\ 112 Chernyshevsky Street, Ufa 450008, Russian Federation
\and {\bf R.I. Yamilov}  \\ 
{\sl E-mail: \url{RvlYamilov@matem.anrb.ru}}\\{\sl URL: \url{http://matem.anrb.ru/en/yamilovri}} \\ Ufa Institute of Mathematics, Russian Academy 
of Sciences, \\ 112 Chernyshevsky Street, Ufa 450008, Russian Federation}

\begin{document}
\maketitle
\abstract{We carry out the generalized symmetry classification of polylinear autonomous discrete equations defined on the square, which belong to a twelve-parametric class. 
The direct result of this classification is a list of equations containing no new examples.
However, as an indirect result of this work we find a number of integrable examples pretending to be new. One of them has a nonstandard symmetry structure, the others are analogues of the Liouville equation in the sense that those are Darboux integrable. We also enumerate all equations of the class, which are linearizable via a two-point first integral, and specify the nature of integrability of some known equations.}

\section{Introduction}

We consider a class of autonomous discrete equations: 
\begin{equation}F(u_{n,m},u_{n+1,m},u_{n,m+1},u_{n+1,m+1})=0, \quad n,m\in \integer,\label{gF}\end{equation} which are analogues of the hyperbolic equations \eq{F(u,u_x,u_y,u_{xy})=0.\label{hyp}} 
In \eqref{gF} $F$ is a polylinear function (a function linear in each of its variables). This class depends on $16$ arbitrary constants. We restrict ourselves to the case when $$\frac{\partial^2 F}{\partial{u_{n+1,m+1}}\partial{u_{n+1,m}}}=0,$$ and the resulting equation depends on $12$ parameters and can be written in the form: 
\eqs{Au_{n+1,m+1}+Bu_{n+1,m}+C=0,\label{ovid}\\
A=a_1u_{n,m}u_{n,m+1}+a_2u_{n,m}+a_3u_{n,m+1}+a_4,\\
B=b_1u_{n,m}u_{n,m+1}+b_2u_{n,m}+b_3u_{n,m+1}+b_4,\\
C=c_1u_{n,m}u_{n,m+1}+c_2u_{n,m}+c_3u_{n,m+1}+c_4.}

The reason we choose this class is technical. The first of the integrability conditions \eqref{zsoh} presented in Theorem \ref{th1} becomes simpler in this case (see a more detailed explanation right after Theorem \ref{th1}).

We use in this paper the generalized symmetry method, see e.g. the review articles \cite{msy87,asy00,y06} for the case of partial differential and difference-differential equations. The generalized symmetry method for discrete equations has been developed in \cite{lpsy08,ly09,mwx11,ly11,ggh11}, and we will use here some results of these works. A review of different methods for testing and classifying discrete equations together with proper references can be found e.g. in \cite{ly11}. 

Following \cite{ly09,ggh11}, we use as an integrability criterion the existence of a non-autono\-mous generalized symmetry of the form:
\begin{equation}\label{i2}
	\frac{d}{dt}u_{n,m} = G_{n,m} (u_{n+1,m}, u_{n-1,m}, u_{n,m}, u_{n,m+1}, u_{n,m-1}).
\end{equation} 
All known integrable discrete equations (\ref{gF}), as well as their non-autonomous generali\-zations \cite{xp09}, possess generalized symmetries of this form, but the only example \cite{a11}. Equations which have such a symmetry (and with no first integrals) are analogues of the sine-Gordone equation $u_{xy} = \sin u$. We are going to enumerate all equations of the sine-Gordon type belonging to the class \eqref{ovid}.

Some of integrable discrete equations of the class \eqref{gF} are called in the literature the discrete KdV and mKdV equations or the discrete sine-Gordon and Liouville equations, see e.g. \cite{sgr11}. From the viewpoint of the generalized symmetry method, such equations are analogues of the hyperbolic equations. As in the hyperbolic case \cite{ZS01, MS11}, the integrable discrete equations have two hierarchies of the generalized symmetries in two different directions and may have first integrals, i.e. may be Darboux integrable (see below). For this reason, we use such terms as sine-Gordon type equations and Liouville type equations taken from the hyperbolic case.

In our approach, the analogues of the Liouville equation $u_{xy} = e^u$ appear in a natural way as equations whose generalized symmetries depend on some arbitrary functions. Such equations are Darboux integrable, i.e. they have two first integrals $W_1,W_2$:
\eq{\label{darb1}(T_1-1)W_2=0,\quad W_2=w_n^{(2)}(u_{n,m+l_2},u_{n,m+l_2+1},\ldots,u_{n,m+k_2}),}
\eq{\label{darb2}(T_2-1)W_1=0,\quad W_1=w_m^{(1)}(u_{n+l_1,m},u_{n+l_1+1,m},\ldots,u_{n+k_1,m}).}
Here $l_1,l_2,k_1,k_2$ are integers, such that $l_1<k_1, \ l_2<k_2$, and $T_1,T_2$ are operators of the shift in the first and second direction, respectively: $T_1 h_{n,m}=h_{n+1,m}$, $T_2 h_{n,m}=h_{n,m+1}$.
We will collect the Darboux integrable equations in a separate list.

In Section \ref{sec2} we present definitions and theoretical statements necessary for this paper. All results are collected in Section \ref{sec3}. Section \ref{sineeq} contains the complete list of sine-Gordon type discrete equations of the class \eqref{ovid}. There are no new integrable examples in that list. New examples appear in the next sections. In Section \ref{bog} an equation with a nonstandard symmetry structure is discussed. Some Darboux integrable equations are collected in Section \ref{darboux}. In Section \ref{line} we enumerate all equations linearizable via a two-point first integral. In Section \ref{addit} we specify the nature of integrability of some known equations. Those equations have been known as equations possessing generalized symmetries. We show that they are either Darboux integrable or equivalent to a linear equation.

 \section{Theory}\label{sec2}

\subsection{Definitions and restrictions}\label{sec21}

Here we are going to discuss some restrictions for discrete equations used in the classification. 

We consider equations of the form \eqref{gF} which are {\it polylinear} and {\it nondegenerate}. The first property means: \seq{\frac{\partial^2 F}{\partial{u_{n,m}^2}}=\frac{\partial^2 F}{\partial{u_{n+1,m}^2}}=\frac{\partial^2 F}{\partial{u_{n,m+1}^2}}=\frac{\partial^2 F}{\partial{u_{n+1,m+1}^2}}=0.} One sometimes uses the following requirement as a nondegeneracy condition: the function $F$ must essentially depend on all its variables. This requirement, however, is not sufficient to exclude equations like the following one: \seq{\tilde F(u_{n,m},u_{n+1,m})\hat F(u_{n,m+1},u_{n+1,m+1})=0.} We rewrite eq. \eqref{gF} in the form \eq{u_{n+1,m+1}=f(u_{n+1,m},u_{n,m},u_{n,m+1})\label{dis}} and require the essential dependence of $f$ on all its variables. So, we have the following nondegeneracy condition in terms of $F$ and $f$: \eq{\label{nondeg}\frac{\partial F}{\partial{u_{n+1,m+1}}},\frac{\partial f}{\partial{u_{n+1,m}}},\frac{\partial f}{\partial{u_{n,m}}},\frac{\partial f}{\partial{u_{n,m+1}}}\neq0.}

We consider {\it generalized symmetries} of the form \eqref{i2} satisfying the conditions
\eq{\label{nondsym}\frac{\partial G_{n,m}}{\partial u_{n+1,m}},\frac{\partial G_{n,m}}{\partial u_{n-1,m}},\frac{\partial G_{n,m}}{\partial u_{n,m+1}},\frac{\partial G_{n,m}}{\partial u_{n,m-1}}\neq0} for all $n,m\in \integer.$ 
These conditions are analogous to ones used in the papers \cite{ly09,ly11} in the autonomous case. 
It can be proved, see \cite{ggh11}, that the function $G_{n,m}$ must have the form
\eq{\label{fun_G}G_{n,m} =\Phi_{n,m} (u_{n+1,m}, u_{n,m}, u_{n-1,m})+\Psi_{n,m} (u_{n,m+1}, u_{n,m}, u_{n,m-1}).}
In practice, for all equations possessing the generalized symmetry \eqref{i2}, we always succeed to find two generalized symmetries of the form
\eq{\frac{d}{d t_1}u_{n,m}=\Phi_{n,m},\quad \frac{d}{d t_2}u_{n,m}=\Psi_{n,m}.\label{two_sym}} Let us recall that the set of symmetries forms a Lie algebra, in particular, \seq{\frac{d}{d\tau_1}u_{n,m}=\alpha \Phi_{n,m}+\beta\Psi_{n,m},\quad \frac{d}{d\tau_2}u_{n,m}=D_{t_1} \Psi_{n,m}-D_{t_2}\Phi_{n,m}} are symmetries too. Here $\alpha,\beta$ are arbitrary constants, $D_{t_1},D_{t_2}$  are operators of total derivative in virtue of eqs. (\ref{two_sym}). The existence of symmetries of the form \eqref{two_sym}
implies the existence of a symmetry given by \eqref{fun_G}. 

The generalized symmetries \eqref{two_sym} as themselves are integrable differential-difference equations. In autonomous case they are integrable equations of the Volterra type of a complete list obtained in  \cite{y83}, see \cite{y06} for details. In non-autonomous case we may get some non-autonomous generalizations of Volterra type equations considered in \cite{ly97}.

It is well known that a function $G_{n,m}$ defining the generalized symmetry \eqref{i2} of eq. \eqref{dis} must satisfy the linearization of this equation \eqref{dis}:
\eq{\label{eq_sym}G_{n+1,m+1}=G_{n+1,m}\frac{\partial f}{\partial{u_{n+1,m}}}+G_{n,m}\frac{\partial f}{\partial{u_{n,n}}}+G_{n,m+1}\frac{\partial f}{\partial{u_{n,m+1}}}.} For fixed values of $n,m$ we can express, using eq. \eqref{dis}, all functions $u_{n+k,m+l}$ ($k,l\neq0$) in terms of the functions \eq{u_{n+k,m},u_{n,m+l},\ \ k,l\in\integer,\label{ind_var}}which can be considered as independent variables. Eq. \eqref{eq_sym} must be identically satisfied for all values of independent variables as well as for any $n,m\in\integer.$

An equation \eqref{dis} is called {\it Darboux integrable} if it has two {\it first integrals} $W_1,W_2$ defined by eqs. (\ref{darb1},\ref{darb2}). We assume that \seq{\frac{\partial W_1}{\partial u_{n+l_1,m}},\frac{\partial W_1}{\partial u_{n+k_1,m}},\frac{\partial W_2}{\partial u_{n,m+l_2}},\frac{\partial W_2}{\partial u_{n,m+k_2}}\neq0} for any $n,m\in\integer$, and $W_i$ is called {\it $N_i$-point} first integral, where $N_i=k_i-l_i+1\geq 2$. As in case of the generalized symmetries, we assume that eqs. (\ref{darb1},\ref{darb2}) are identically satisfied for all values of \eqref{ind_var} and for all $n,m\in\integer.$ In other words, eqs. (\ref{darb1},\ref{darb2}) must be identically satisfied on solutions of the corresponding discrete equation.

It is not difficult to prove that the function $W_1$ cannot depend on independent variables $u_{n,m+l},l\neq0,$ and the function $W_2$ cannot depend on variables $u_{n+k,m},k\neq0$. It is obvious that, for any first integrals $W_1,W_2$, arbitrary functions $\Omega_1,\Omega_2$ of the form \eqs{\label{arb_fun}\Omega_1=\Omega_1(n,W_1,T_1^{\pm1}W_1,\ldots,T_1^{\pm j_1}W_1),\\ \Omega_2=\Omega_2(m,W_2,T_2^{\pm1}W_2,\ldots,T_2^{\pm j_2}W_2)} are also the first integrals.

Darboux integrable equations are transformed into linear equations for  $W_1,W_2$ shown in (\ref{darb1},\ref{darb2}), whose solutions are obvious. So, in this case, solutions of eq. \eqref{dis} may be found from ordinary discrete equations: \eq{\label{zam_lin}W_1=\alpha_n,\quad W_2=\beta_m.} Here $\alpha_n,\beta_m$ are arbitrary functions of integration which, in the discrete case, are nothing but the sets of a priori independent arbitrary constants:\seq{\{\alpha_0,\alpha_{\pm 1},\alpha_{\pm 2},\ldots\},\quad \{\beta_0,\beta_{\pm 1},\beta_{\pm 2},\ldots\}.} One often can find for a Darboux integrable equation a more simple linearizing trans\-formation than ones shown in (\ref{darb1}) and (\ref{darb2}), which may be more convenient for construct\-ing solutions. Such transformations are not discussed in the present paper.

The sine-Gordon equation has the generalized symmetries, but does not have any first integrals. It is difficult to prove that an equation has no first integrals. We will search for equations of the sine-Gordon type, but the second property will be checked for a fixed and low number of points $2\leq N\leq 5$. The upper bound $5$ is the number for which we are able to solve the problem at the moment.

In our classification we are going to reject the linear equations as well as equations equivalent to the linear ones. The following equations are transformed into some linear ones by the point transform $u_{n,m}=e^{v_{n,m}}$: 
\eqs{
  u_{n+1,m+1}u_{n,m}u_{n,m+1}=\nu u_{n+1,m}\label{l1},\\
  u_{n+1,m+1}u_{n,m}=\nu u_{n+1,m}u_{n,m+1},\\	
  u_{n+1,m+1}u_{n,m+1}=\nu u_{n+1,m}u_{n,m},\\	
  u_{n+1,m+1}=\nu u_{n+1,m}u_{n,m}u_{n,m+1},}
where $\nu \ne 0$ is an arbitrary constant. We reject equations related to eqs. \eqref{l1} or to a linear equation by an autonomous M\"obius (linear-fractional) transformation: \eq{\hat u_{n,m}=\frac{\alpha u_{n,m}+\beta}{\gamma u_{n,m}+\delta}\label{mob}.} While, in  general, the class \eqref{ovid} is not invariant under the M\"obius transformations, sometimes a transformation \eqref{mob} may leave an equation within the class \eqref{ovid}.

In our classification, we will search for equations of the form \eqref{ovid} which have the following properties:

\begin{Spisok} {\bf (properties used in the classification)}\label{prop}
\begin{enumerate}
\item Eq. \eqref{ovid} is nondegenerate, i.e. it satisfies the conditions \eqref{nondeg};
\item It possesses a generalized symmetry of the form (\ref{i2},\ref{nondsym});
\item It does not have any $N$-point first integrals with $2\leq N\leq5$; 
\item Eq. \eqref{ovid} is nonlinear and is not equivalent in the sense of \eqref{mob} to any of eqs. \eqref{l1}. 
\end{enumerate}
\end{Spisok}

\subsection{Method of classification}

Let us recall that a conservation law of eq. \eqref{dis} is a relation of the form $(T_1-1)p_{n,m}=(T_2-1)q_{n,m}$ which holds on the solutions of this equation. Here $p_{n,m},q_{n,m}$ may depend explicitly on the discrete variables $n,m$ and on a finite number of the functions $u_{n+k,m+l}$. We will use below four integrability conditions which have the form of conservation laws:
\eq{(T_1-1)p_{n,m}^{(j)}=(T_2-1)q_{n,m}^{(j)}, \quad j=1,2,3,4.\label{zsoh}}
The following statement takes place:

\begin{Theorem}\label{th1} If an equation \eqref{dis} has a generalized symmetry of the form (\ref{i2},\ref{nondsym}), then it must have conservation laws \eqref{zsoh}, such that:
\eqs{\label{a10}
 p^{(1)}_{n,m} = \log f_{u_{n+1,m}} , \qquad
 q^{(1)}_{n,m} = q^{(1)}_{n,m} (u_{n+2,m},u_{n+1,m},u_{n,m}) ;\\
 p^{(2)}_{n,m} = \log \frac{f_{u_{n,m}}}{f_{u_{n,m+1}}} , \qquad
 q^{(2)}_{n,m} = q^{(2)}_{n,m} (u_{n+2,m},u_{n+1,m},u_{n,m}) ;\\
 q^{(3)}_{n,m} = \log f_{u_{n,m+1}} , \qquad
 p^{(3)}_{n,m} = p^{(3)}_{n,m} (u_{n,m+2},u_{n,m+1},u_{n,m}) ;\\
 q^{(4)}_{n,m} = \log \frac{f_{u_{n,m}}}{f_{u_{n+1,m}}} , \qquad
 p^{(4)}_{n,m} = p^{(4)}_{n,m} (u_{n,m+2},u_{n,m+1},u_{n,m}) ,
}where $f_{u_{n+k,m+l}}=\frac{\partial f}{\partial u_{n+k,m+l}}.$
\end{Theorem}

Here the functions $p^{(1)}_{n,m},p^{(2)}_{n,m},q^{(3)}_{n,m},q^{(4)}_{n,m}$  are defined explicitly in terms of eq.  \eqref{dis}. The other four functions must be found from eqs. \eqref{zsoh} and must have the form shown in \eqref{a10}. We do not give the proof, as this theorem has been taken from \cite{ly09}, where the completely autonomous case is   considered. In \cite{ggh11} the same result is presented for the partially non-autonomous case we consider here. In this paper we just give more precise definitions and formulations suitable for the paper.

It should be remark that in case of the class \eqref{ovid} the function $p^{(1)}_{n,m}$ depends on two variables $u_{n,m},\ u_{n,m+1}$ only, while in the general case \eqref{gF} it may depend on $u_{n+1,m}$ too. This property makes the calculation simpler and explains our choice of the class \eqref{ovid}.

The conservation laws (\ref{zsoh},\ref{a10}) follow from eq. \eqref{eq_sym}, and the unknown functions $q^{(1)}_{n,m},q^{(2)}_{n,m},p^{(3)}_{n,m},p^{(4)}_{n,m}$ are related to $G_{n,m}$ of eq. \eqref{i2} by the following formulae:
\eqs{q^{(1)}_{n,m}=-T_1\log\frac{\partial G_{n,m}}{\partial u_{n+1,m}},\quad q^{(2)}_{n,m}=T_1\log\frac{\partial G_{n,m}}{\partial u_{n-1,m}};\\
p^{(3)}_{n,m}=-T_2\log\frac{\partial G_{n,m}}{\partial u_{n,m+1}},\quad p^{(4)}_{n,m}=T_2\log\frac{\partial G_{n,m}}{\partial u_{n,m-1}}.\label{dif_sym}}
If for a given discrete equation the integrability conditions (\ref{zsoh},\ref{a10}) are satisfied, i.e. there exist some functions $q^{(1)}_{n,m},q^{(2)}_{n,m},p^{(3)}_{n,m},p^{(4)}_{n,m}$ of the form shown in \eqref{a10}, then we can construct, in principle, a generalized symmetry by using eqs. \eqref{dif_sym} (see \cite{ly11} for a detailed explanation).

Our aim is to find a function $f$ defining eq. \eqref{dis}, for which the equations (\ref{zsoh},\ref{a10}) are solvable. The problem is, however, that eqs. (\ref{zsoh},\ref{a10}) are functional-difference equations. We solve them, reducing to a system of linear partial differential equations for the unknown functions $q^{(1)}_{n,m},q^{(2)}_{n,m},p^{(3)}_{n,m},p^{(4)}_{n,m}$, and we do that by using some annihilation operators (or {\it annihilators}). Conditions of solvability of such a system give an algebraic system of equations for coefficients of eq. \eqref{ovid}. So, the classification problem is reduced to solving such algebraic systems.
 
The annihilators are defined as: \eqs{Y_{k}=T_2^{-k}\frac{\partial}{\partial u_{n,m+1}}T_2^k,\quad Y_{-k}=T_2^k\frac{\partial}{\partial u_{n,m-1}}T_2^{-k},\quad k>0;\\\label{anih} Z_{k}=T_1^{-k}\frac{\partial}{\partial u_{n+1,m}}T_1^k,\quad Z_{-k}=T_1^k\frac{\partial}{\partial u_{n-1,m}}T_1^{-k},\quad k>0.} These operators are the differentiation operators, and their action will be clarified by example of $Y_1$.

It is obvious that
\seq{Y_1 u_{n,m+l}=0,\quad l\neq0;\qquad Y_1 u_{n,m}=1.} On the other independent variables \eqref{ind_var} defined at a fixed point $(n,m)$, it acts as follows: \seqs{Y_1 u_{n+k,m}=\prod_{j=0}^{k-1}T_2^{-1}T_1^j\frac{\partial f^{(1,1)}}{\partial u_{n,m+1}},\quad k>0;\\
Y_1 u_{n+k,m}=\prod_{j=k+1}^{0}T_2^{-1}T_1^{j}\frac{\partial f^{(-1,1)}}{\partial u_{n,m+1}},\quad k<0.} Here we denote: \eq{\label{4forms}u_{n+k,m+l}=f^{(k,l)}(u_{n+k,m},u_{n,m},u_{n,m+l}),\quad k,l\in\{-1,1\}.}In particular $f=f^{(1,1)}$, and we have in \eqref{4forms} four equivalent and obvious forms of eq. \eqref{dis}. So, for functions depending on \eqref{ind_var}, we have:
\seqs{
Y_1 =\sum_{j=-\infty}^{\infty}Y_1 (u_{n+j,m})\frac{\partial }{\partial u_{n+j,m}}.} 

In case of the other operators $Y_{k},Z_{k},k\neq 0$, as well as for a standard differentiation operator $L$, we find at first the results $L(u_{n+k,m}),\ L(u_{n,m+l})$ of its action on the independent variables \eqref{ind_var} by using the definition \eqref{anih}. Then we can use the general property of the differentiation operators in order to apply $L$ to an arbitrary function of the independent variables:
\seq{L=\sum_{k=-\infty}^\infty L(u_{n+k,m})\frac{\partial}{\partial u_{n+k,m}}+\sum_{l=-\infty,l\neq0}^\infty L(u_{n,m+l})\frac{\partial}{\partial u_{n,m+l}}. }
It is clear that, acting on functions which depend on a finite number of variables, we obtain the finite expressions, for instance, \seqs{Y_{-1}h(u_{n+1,m},u_{n,m},u_{n-1,m})=&T_2\left(\frac{\partial f^{(1,-1)}}{\partial u_{n,m-1}}\right)\frac{\partial h}{\partial u_{n+1,m}}+\frac{\partial h}{\partial u_{n,m}}\\&+T_2\left(\frac{\partial f^{(-1,-1)}}{\partial u_{n,m-1}}\right)\frac{\partial h}{\partial u_{n-1,m}}.}

The operators \eqref{anih} have been introduced in \cite{h05} and applied there to the first integrals. Particular cases of these operators have been applied to the conservation laws in \cite{rh07} and to the autonomous integrability conditions (\ref{zsoh},\ref{a10}) in \cite{ly11}. In \cite{ggh11} these operators have been used in the study of the partially non-autonomous integrability conditions (\ref{zsoh},\ref{a10}) and of their generalizations.

\begin{Lemma} \label{lemanih}The operators \eqref{anih} annihilate the following functions: \seqs{Y_{-l}q^{(j)}_{n+k,m+l}=0,\quad j=1,2,}for all $l\neq0$ and $k$;\seqs{Z_{-k}p^{(j)}_{n+k,m+l}=0,\quad j=3,4,}for all $k\neq0$ and $l$.\end{Lemma} 

\paragraph{Proof.} We give the proof in one of the cases, as the others are quite similar:
\seqs{Y_{-l}&q^{(1)}_{n+k,m+l}=T_2^l\frac{\partial}{\partial u_{n,m-1}}T_2^{-l} T_2^l q^{(1)}_{n+k,m}\\
=&T_2^l\frac{\partial}{\partial u_{n,m-1}} q^{(1)}_{n+k,m}(u_{n+2+k,m},u_{n+1+k,m},u_{n+k,m})=0,}
where $l > 0$. \qed

\paragraph{The use of annihilators.}
When applying the operators
\seqs{\Lambda_{2,l}=\left\{\begin{split}T_1^{-1}\sum_{j=0}^{l-1}T_2^j,\quad l>0,\\ -T_1^{-1}\sum_{j=l}^{-1}T_2^j,\quad l<0,\end{split}\right.\quad \Lambda_{1,k}=\left\{\begin{split}T_2^{-1}\sum_{j=0}^{k-1}T_1^j,\quad k>0,\\ -T_2^{-1}\sum_{j=k}^{-1}T_1^j,\quad k<0,\end{split}\right.} to the relations \eqref{zsoh}, we get: \eqs{\Lambda_{2,l}(T_1-1)p^{(j)}_{n,m}=(T_2^l-1)q^{(j)}_{n-1,m},\quad l\neq0,\quad j=1,2; \label{soh2}\\
(T_1^k-1)p^{(j)}_{n,m-1}=\Lambda_{1,k}(T_2-1)q^{(j)}_{n,m},\quad k\neq0,\quad j=3,4.}
Then, applying $Y_{-l},\ Z_{-k}$ to eqs. \eqref{soh2}, respectively, and taking into account Lemma \ref{lemanih}, we are led to:\spisok{\label{syspde1}Y_{-l}q^{(j)}_{n-1,m}=r^{(l,j)}_{n,m},\quad l\neq0,\quad j=1,2;\\\label{syspde2} Z_{-k}p^{(j)}_{n,m-1}=s^{(k,j)}_{n,m},\quad k\neq0,\quad j=3,4.} Here the functions $r^{(l,j)}_{n,m},s^{(k,j)}_{n,m}$ as well as coefficients of $Y_{-l},Z_{-k}$ are explicitly expressed in terms of eq. \eqref{ovid}.

We have an infinite system of linear non-homogeneous PDEs for each of four functions $q^{(j)}_{n-1,m},\ p^{(j)}_{n,m-1}.$ We can add to eqs. (\ref{syspde1},\ref{syspde2}) linear PDEs with the commuta\-tors $[Y_{l_1},Y_{l_2}]$ and $[Z_{k_1},Z_{k_2}]$ instead of the operators $Y_{-l}$ and $Z_{-k}$. Also we can decom\-pose each equation with respect to additional variables $u_{n,m+j}$ in the case of \eqref{syspde1} and with respect to $u_{n+j,m}$ in the case of \eqref{syspde2}. 

Each of the unknown functions depends on three variables, so any four equations for each of the unknown functions must be linear dependent. This provides the necessary conditions of integrability which are expressed in terms of the coefficients of eq. \eqref{ovid} only. In practice, eqs. (\ref{syspde1},\ref{syspde2}) with $l,k\in\{\pm1,\pm2\}$ are enough for the classification of eqs. \eqref{ovid}.

\paragraph{First integrals.}
Eqs. (\ref{darb1},\ref{darb2}) for the first integrals imply the following relations:
\eq{\label{fi2}T_1^kW_2=W_2,\qquad T_2^l W_1=W_1,} with arbitrary integer $k,l.$ The operators $Z_{-k},\ Y_{-l}$ annihilate the left hand sides of eqs. \eqref{fi2}, respectively. We get infinite systems of linear homogeneous PDEs for the first integrals $W_1,W_2:$ \eq{Z_{-k}W_2=0,\ k\neq0,\qquad Y_{-l}W_1=0,\ l\neq0.\label{sysin}}

By using the systems of equations \eqref{sysin}, we can solve the classification problem for equations like \eqref{ovid}, see Section \ref{line}. We can also find the first integrals for a given equation, see Section \ref{darboux}. Directly from the systems \eqref{sysin}, we can find a dependence of first integrals on the independent variables \eqref{ind_var}, an explicit dependence on $n,m$ is specified as it is shown in Section \ref{line}.

\section{Results}\label{sec3}

\subsection{Classification theorem}\label{sineeq}

As a result of the generalized symmetry classification, we are led to the following list of discrete equations:

\begin{Spisok} {\bf (eqs. \eqref{ovid} of the sine-Gordon and Burgers type together with their symmetries of the form \eqref{two_sym})}\label{sineql}
\begin{enumerate}
\item
\seqs{(u_{n+1,m+1}-1)(u_{n,m+1}+1)=(u_{n+1,m}+1)(u_{n,m}-1),\\
\frac{d}{dt_1}u_{n,m}=(u_{n,m}^2-1)(u_{n+1,m}-u_{n-1,m}),\\
\frac{d}{dt_2}u_{n,m}=(u_{n,m}^2-1)\left(\frac1{u_{n,m+1}+u_{n,m}}-\frac1{u_{n,m}+u_{n,m-1}}\right)}

\item
\seqs{(u_{n+1,m+1}-u_{n+1,m}+c_2)(u_{n,m}-u_{n,m+1}-c_2)+u_{n+1,m}-u_{n,m+1}+c_4=0,\\
\frac{d}{dt_1}u_{n,m}=(u_{n+1,m}-u_{n,m}+c_2+c_4)(u_{n,m}-u_{n-1,m}+c_2+c_4),\\
\frac{d}{dt_2}u_{n,m}=\frac{(u_{n,m+1}-u_{n,m})(u_{n,m}-u_{n,m-1})-c_2-c_2^2}{u_{n,m+1}-u_{n,m-1}+2c_2+1}}

\item 
\seqs{u_{n+1,m+1}u_{n,m+1}+b_2u_{n+1,m}u_{n,m}+c_1u_{n,m}u_{n,m+1}=0,\quad b_2,c_1\neq0,\\
\frac{d}{dt_1}u_{n,m}=\frac{u_{n+1,m}u_{n,m}}{u_{n-1,m}},\qquad
\frac{d}{dt_2}u_{n,m}=\frac{u_{n,m}u_{n,m-1}}{u_{n,m+1}-b_2u_{n,m-1}}}

\item
\seqs{u_{n+1,m+1}u_{n,m}+u_{n+1,m}u_{n,m+1}+a_3u_{n+1,m+1}u_{n,m+1}+b_2u_{n+1,m}u_{n,m}\\+c_1u_{n,m}u_{n,m+1}=0,\quad
c_1,b_2\neq0,\\
\frac{d}{dt_1}u_{n,m}=(1-a_3b_2)\frac{u_{n+1,m}u_{n,m}}{u_{n-1,m}}+c_1\left(u_{n+1,m}+\frac{u_{n,m}^2}{u_{n-1,m}}\right),\\
\frac{d}{dt_2}u_{n,m}=\frac{(u_{n,m+1}+b_2u_{n,m})(u_{n,m}+b_2u_{n,m-1})}{a_3u_{n,m+1}-b_2u_{n,m-1}}}

\item
\seqs{u_{n+1,m+1}u_{n,m}+u_{n+1,m}u_{n,m+1}+a_3u_{n+1,m+1}u_{n,m+1}\\ +c_1u_{n,m}u_{n,m+1}=0,\quad c_1,a_3\neq0,\\
\frac{d}{dt_1}u_{n,m}=\frac{u_{n+1,m}u_{n,m}}{u_{n-1,m}}+c_1\left(u_{n+1,m}+\frac{u_{n,m}^2}{u_{n-1,m}}\right),\\
\frac{d}{dt_2}u_{n,m}=\frac{u_{n,m-1}u_{n,m}}{u_{n,m+1}}+a_3\left(u_{n,m-1}+\frac{u_{n,m}^2}{u_{n,m+1}}\right)}

\item Discrete Burgers equation:
\spisok{(u_{n+1,m+1}+d)(u_{n,m}+a_3)u_{n,m+1}+b_1(u_{n,m+1}+d)(u_{n+1,m}+a_3)u_{n,m}=0,\label{bur}\\
a_3,b_1,d,|d+b_1a_3|+|b_1^2-1|\neq0,\nonumber\\
\frac{d}{dt_1}u_{n,m}=(-b_1)^{-m}u_{n,m}(u_{n+1,m}-u_{n,m})C_1+(-b_1)^{m}\frac{u_{n-1,m}-u_{n,m}}{u_{n-1,m}}C_2 , \label{bur_s1}\\
\frac{d}{dt_2}u_{n,m}=(-b_1)^{n}\frac{(u_{n,m}-u_{n,m+1})(u_{n,m}+a_3)}{u_{n,m+1}+d}C_3\nonumber\\+(-b_1)^{-n}\frac{(u_{n,m}-u_{n,m-1})(u_{n,m}+d)}{u_{n,m-1}+a_3}C_4\label{bur_s2}}

\end{enumerate}
\end{Spisok}

\begin{Theorem}\label{th_clas} An equation of the form \eqref{ovid} possesses properties enumerated in List \ref{prop} if and only if it is equivalent to an equation of List \ref{sineql} up to a linear transformation $\hat u_{n,m}=\alpha u_{n,m}+\beta,$ where $\alpha \ne 0$ and $\beta$ are constants. For eq. 6 of List \ref{sineql}, the property 3 of List \ref{prop} has been checked for $N=2,3$.
\end{Theorem}

\paragraph{\small Remark.} More precisely, we search for equations possessing properties of List \ref{prop} and obtain List \ref{sineql}. The resulting equations have the properties 1,2,4. The property 3 has been checked for all equations of List \ref{sineql} except for eq. 6. In this case we check the property only for $N=2,3$ because of computational complexity.

Really, there are some special cases of eq. 6 with more complicated first integrals. An exhaustive investigation of this problem is left for a future work. 

As for eqs. 1-5, we hope that these equations do not have any first integrals at all, however, checking this property is an open problem which is also left for a future work.
\bigskip

All equations of List \ref{sineql} are known. Eqs. 1-5 can be found in the paper \cite{ly11}, most of them in a slightly different form. Sometimes, transformations of the form \eq{\hat u_{n,m}=\frac{\alpha_{n,m}u_{n,m}+\beta_{n,m}}{\gamma_{n,m}u_{n,m}+\delta_{n,m}},\label{MOBnm}} i.e. non-autonomous M\"obius transformations, as well as the transformations $n\leftrightarrow m$ and $m\rightarrow -m$ leave the equations within the class \eqref{ovid}. 
For example, the transform $$\hat u_{n,m}=u_{n,m}+\alpha n+\beta m$$ changes in eq. 2 of List \ref{sineql} the coefficients $c_2,c_4$ only, and we can get any values of $c_2,c_4$, in particular $c_2=-1/2,c_4=0$. The form of eqs. 3-5 of List \ref{sineql} is invariant under the transform \eq{ \hat u_{n,m}=u_{n,m}\alpha^n\beta^m\label{zam_mn}.} In the case of eq. 3 we can make $b_2=c_1=1$. The same transformation \eqref{zam_mn} allows us to obtain $a_3=0,b_2=c_1=1$ or $a_3=b_2=c_1\neq0$ in eq. 4 and $a_3=c_1=1$ in eq. 5. All the particular cases of eqs. 2-5, mentioned above, as well as eq. 1 are presented in \cite{ly11}. As it has been noticed in \cite{ly11}, the equations 4 with $a_3=0$ and 5 are equivalent to each other up to the transform $m\rightarrow -m$.

A detailed discussion of eqs. 1-5 of List \ref{sineql} and of their generalized symmetries together with proper references are presented in the paper \cite{ly11}. We just remark here that most of these equations, namely eqs. 2-5, come from \cite{hv07}, where they are given in a more general form. In \cite{ly11} the authors have selected from those more general equations all sine-Gordon type particular cases which are out of the well-known Adler-Bobenko-Suris list \cite{abs03}.

While the equations 1-5 of the above list are pure analogues of the sine-Gordon equation, eq. 6 is a Burgers type equation. The generalized symmetries (\ref{bur_s1},\ref{bur_s2}) depend on arbitrary constants $C_1,C_2,C_3,C_4$ which may be equal to zero, and this is natural for Burgers type equations. As well as in the continuous case, eq. 6 is derived from a linear equation by a discrete analogue of the Hopf-Cole transform: 
\eq{u_{n,m}=\frac{v_{n+1,m}}{v_{n,m}}.\label{bur_zam}}

The following non-autonomous linear equation: \eq{v_{n+1,m+1}=\alpha_{n,m}v_{n+1,m}+\beta_{n,m}v_{n,m+1}+\gamma_{n,m}v_{n,m},\label{bur_lin}} where $\alpha_{n,m},\beta_{n,m},\gamma_{n,m}\neq0$ for any values of $n,m$, is transformed by \eqref{bur_zam} into \eqs{(u_{n+1,m+1}-\beta_{n+1,m})(\alpha_{n,m}u_{n,m}+\gamma_{n,m})u_{n,m+1}\\=(u_{n,m+1}-\beta_{n,m})(\alpha_{n+1,m}u_{n+1,m}+\gamma_{n+1,m})u_{n,m}.\label{ur_bur_o}} The equation \eqref{bur} is a particular case of eq. \eqref{ur_bur_o} corresponding to \eq{\alpha_{n,m}=(-b_1)^n,\quad \beta_{n,m}=-d,\quad \gamma_{n,m}=a_3(-b_1)^n\label{bur_konk}.}
The generalized symmetries (\ref{bur_s1},\ref{bur_s2}) are derived from the symmetries  \seqs{\frac{d}{dt_1}v_{n,m}=C_1(-b_1)^{-m}v_{n+1,m}+C_2(-b_1)^mv_{n-1,m},\\ \frac{d}{dt_2}v_{n,m}=-C_3v_{n,m+1}-C_4v_{n,m-1}} of eq. (\ref{bur_lin},\ref{bur_konk}) by the same transformation \eqref{bur_zam}.

Eq. \eqref{bur} in the case $b_1=-1$ has been found in \cite{h04}. The general form  \eqref{ur_bur_o} together with the transformation \eqref{bur_zam} are presented in \cite{rj06}. 
We have here a slight autonomous generalization of eq. \eqref{bur} with $b_1=-1$ which, however, has the non-autonomous generalized symmetries and which is related to a non-autonomous linear equation.

It is interesting that there are the Darboux integrable equations among such Burgers type equations.
A non-degenerate equation of the form \eqref{bur} has two $N$-point first integrals with $2\leq N\leq3$ if and only if 
\eq{\label{bur_dar}a_3,b_1,d\neq0,\ \ d+b_1a_3=b_1^2-1=0.}  In the case $b_1=1$, eq. (\ref{bur},\ref{bur_dar}) is equivalent to eq. 6 of List \ref{Darboux1} (see below) up to a rescale of $u_{n,m}$. In the case $b_1=-1$ we obtain eq. 4 of List \ref{Darboux2} by using a M\"obius transformation \eqref{mob} and the transformation $n\leftrightarrow m$.

\subsection{Example with a nonstandard symmetry structure}\label{bog}

Among integrable hyperbolic equations of the form \eqref{hyp}, there are not only the sine-Gordon and Louville equations but also the Tzitzeica equation \eq{u_{xy}=e^u+e^{-2u},} see e.g. \cite{t907,zs79}. This equation is not Darboux integrable and it differs from the sine-Gordon equation in a more complicated structure of the generalized symmetries.
Here we present an equation whose generalized symmetries have a more complicated structure than symmetries presented in List \ref{sineql}.

The equation
\eq{ u_{n+1,m+1}(u_{n,m}-u_{n,m+1})-u_{n+1,m}(u_{n,m}+u_{n,m+1})+1=0\label{tzit}}
satisfies all the integrability conditions (\ref{zsoh},\ref{a10}) and does not have any $N$-point first integral, such that $2\leq N\leq7$. It has one of generalized symmetries of the form \eqref{two_sym}:
\eq{\frac{d}{dt_2}u_{n,m}=(-1)^n\frac{u_{n,m+1}u_{n,m-1}+u_{n,m}^2}{u_{n,m+1}+u_{n,m-1}}.\label{s1_tz}}
However, there is no second symmetry of the form \eqref{two_sym} in this case as well as no five-point symmetry (\ref{i2},\ref{nondsym}). 

We have found a more complicated generalized symmetry in the $n$-direction:
\eq{\label{s2_tz}\frac{d}{dt_1}u_{n,m}=h_{n,m}h_{n-1,m}(a_nu_{n+2,m} - a_{n-1}u_{n-2,m}),} where
\[h_{n,m}=1-2u_{n+1,m}u_{n,m},\quad a_{n+2}=a_{n}.\] We have done that by using a technique developed in \cite{ggh11}. The symmetry depends on an arbitrary two-periodic function $a_n$ which can be expressed in the form: \eqs{a_n=\tilde a+\hat{a}(-1)^n.} 
There are here the autonomous particular case $a_n=1$ and the non-autonomous one $a_n=(-1)^n$, and they generate all the other possible subcases as linear combinations. For instance, we have
\eqs{a_n=\frac{1+(-1)^n}2=\left\{\begin{array}{l} 0,\quad n=2k+1;\\1,\quad n=2k.\end{array}\right.}

The generalized symmetries (\ref{s1_tz},\ref{s2_tz}) in themselves exemplify integrable differential-difference equations. One can see that $n$ is an outer parameter in eq. \eqref{s1_tz}, and this equation is really a known 1+1-dimensional autonomous equation of the Volterra type \cite{y06}. In the case of eq. \eqref{s2_tz}, we have the essentially non-autonomous differential-difference equation with two-periodic coefficient $a_n$.

Eq. \eqref{s2_tz} possesses the following two conservation laws:
\[\frac{d}{dt_1}p_{n,m}^{(i)}=(T_1-1)q^{(i)}_{n,m},\ i=1,2,\] with the conserved densities
\[p^{(1)}_{n,m}=\log h_{n,m},\quad p^{(2)}_{n,m}=a_nh_{n+1,m}h_{n-1,m}-a_{n-1}h_{n,m} 
-2a_nu_{n+2,m}u_{n-1,m}.\]
The functions $q^{(i)}_{n,m}$ are found automatically, see e.g. \cite{y06}.
There is the following generalized symmetry of eq. \eqref{s2_tz}:
\eqs{\label{s3_tz}\frac{d}{dt'_1}u_{n,m}=h_{n,m}h_{n-1,m}(
  b_nu_{n+4,m}h_{n+2,m}h_{n+1,m} - b_{n-1}h_{n-2,m}h_{n-3,m}u_{n-4,m}\\ 
+ 2u_{n,m}( b_{n-1}u_{n+3,m}u_{n-2,m}h_{n+1,m} - b_nh_{n-2,m}u_{n+2,m}u_{n-3,m} )\\ 
+ 2( 2u_{n+1,m}u_{n,m}u_{n-1,m} - u_{n+1,m} - u_{n-1,m} )(b_nu_{n+2,m}^2 - b_{n-1}u_{n-2,m}^2)\\
+ 2u_{n,m}( b_{n-1}u_{n+1,m}u_{n-2,m} - b_nu_{n+2,m}u_{n-1,m} )),\quad b_{n+2}=b_n.}
It can be checked that eq. \eqref{s3_tz} is a generalized symmetry of the discrete equation \eqref{tzit} too. It can also be checked that any two symmetries \eqs{D_{t'}u_{n,m}=G^{'}_{n,m},\quad  D_{t''}u_{n,m}=G^{''}_{n,m}} of the form \eqref{s2_tz} or \eqref{s3_tz} commute: $$[D_{t'},D_{t''}]u_{n,m}=D_{t'}G^{''}_{n,m}-D_{t''}G^{'}_{n,m}=0.$$

Among the integrable differential-difference equations of the form
\eqs{\frac{\partial }{\partial \tau }v_n=H(v_{n+2},v_{n+1},v_n,v_{n-1},v_{n-2}),} 
there exist not only generalized symmetries of Volterra type equations but also the lowest terms of their own hierarchies. Equations of the second type may be called the Itoh-Narita-Bogoyavlensky equations, see e.g.  \cite{bo88,na82,i75}, for example \eqs{\partial_\tau v_n=v_n(v_{n+2}+v_{n+1}-v_{n-1}-v_{n-2}).} The most recent examples of this type can be found in \cite{a11,s11}, and all such examples are autonomous. Up to our knowledge, eq. \eqref{s2_tz} is a new integrable example of this kind, moreover, it is non-autonomous. The discrete equation \eqref{tzit} is possibly new example similar to the Tzitzeica equation. Another equation like that is discussed in \cite{a11}.

Let us notice that there is one more integrable equation of the class \eqref{ovid} which has the same properties as eq. \eqref{tzit}:
\seq{u_{n+1,m+1}(u_{n,m}+u_{n,m+1})+u_{n+1,m}(u_{n,m}-u_{n,m+1})-1=0.}
This example is, however, trivial in the sense that it is related to eq. \eqref{tzit} by the transformation $\hat u_{n,m}=(-1)^{m}u_{n,m}$. 

\subsection{Darboux integrable equations}\label{darboux}

Classifying the discrete equations \eqref{ovid}, we find the cases when the conservation laws (\ref{zsoh},\ref{a10}) and generalized symmetries \eqref{two_sym} depend on arbitrary functions of the form \eqref{arb_fun}. Such situation is explained by the existence of first integrals. 

In cases like that the systems of PDEs (\ref{syspde1},\ref{syspde2}), we use in the symmetry classifica\-tion, are degenerate, and their solutions also depend on some arbitrary functions. In all such cases we verify whether a discrete equation is Darboux integrable. If the result is positive, we include the equation in one of two lists given below. 

List \ref{Darboux1} contains equations which have $N$-point first integrals, such that $N\geq3$. Any equation presented in List \ref{Darboux2} has one 2-point first integral.

\begin{Spisok}\label{Darboux1}{\bf (Darboux integrable equations with their first integrals $W_1,W_2$: both $N$-point integrals are such that $N\geq3$)}
\begin{enumerate}
\item 
\[(u_{n+1,m+1}-u_{n+1,m})(u_{n,m}-u_{n,m+1})+u_{n+1,m+1}+u_{n+1,m}+u_{n,m+1}+u_{n,m}=0,\]
\[W_1=\frac{2(u_{n+1,m}+u_{n,m})+1}{(u_{n+2,m}-u_{n,m})(u_{n+1,m}-u_{n-1,m})},\]
\[W_2=(-1)^n\frac{u_{n,m+1}+u_{n,m-1}-2(u_{n,m}+1)}{u_{n,m+1}-u_{n,m-1}}\]

\item 
\[u_{n+1,m+1}(u_{n,m}+b_2u_{n,m+1})+u_{n+1,m}(b_2u_{n,m}+u_{n,m+1})+c_4=0,\quad b_2,|b_2^2-1|+|c_4|\neq0,\]
\[W_1=\frac{u_{n+1,m}u_{n,m}(b_2^2-1)+b_2c_4}{(u_{n+2,m}-u_{n,m})(u_{n+1,m}-u_{n-1,m})},\]
\[W_2=(-1)^n\frac{b_2(u_{n,m+1}+u_{n,m-1})+2u_{n,m}}{u_{n,m+1}-u_{n,m-1}}\]

\item 
\[(u_{n+1,m}+a_3u_{n+1,m+1})(u_{n,m}+a_3u_{n,m+1})+u_{n+1,m}+u_{n,m}+\frac1{a_3+1}=0,\quad a_3\neq-1,0,\]
\[W_1=(-a_3)^{-m}\frac{(a_3+1)(u_{n+1,m}+u_{n,m})+1}{(u_{n+2,m}-u_{n,m})(u_{n+1,m}-u_{n-1,m})},\]
\[W_2=\left(\frac{u_{n,m-1}+a_3u_{n,m}+1}{\sqrt{-a_3}(u_{n,m}+a_3u_{n,m+1})}\right)^{(-1)^n}\]

\item 
\[u_{n+1,m+1}u_{n,m}+b_3u_{n+1,m}u_{n,m+1}+1=0,\ \ b_3=\pm1,\]
\[W_1=(-b_3)^m\frac{u_{n+1,m}-b_3u_{n-1,m}}{u_{n,m}},\]
\[W_2=(-b_3)^n\frac{u_{n,m+1}-b_3u_{n,m-1}}{u_{n,m}}\]

\item 
\[(u_{n+1,m+1}+1)(u_{n,m}+1)-(u_{n+1,m}-1)(u_{n,m+1}-1)=0,\]
\[W_1=\frac{u_{n,m}^2-1}{(u_{n+1,m}+u_{n,m})(u_{n-1,m}+u_{n,m})},\]
\[W_2=\frac{u_{n,m}^2-1}{(u_{n,m+1}+u_{n,m})(u_{n,m-1}+u_{n,m})}\]

\item 
\[(u_{n+1,m+1}-1)(u_{n,m}+1)u_{n,m+1}+(u_{n,m+1}-1)(u_{n+1,m}+1)u_{n,m}=0,\]
\[W_1=(-1)^m\frac{(u_{n+1,m}u_{n,m}-1)u_{n-1,m}}{u_{n-1,m}u_{n,m}-1},\]
\[W_2=(-1)^n\frac{(u_{n,m}+u_{n,m+1})(u_{n,m-1}+1)}{(u_{n,m+1}-1)(u_{n,m-1}+u_{n,m})}\]
\end{enumerate}
\end{Spisok}
\bigskip

Up to our knowledge, the non-autonomous first integrals have not been considered in the literature. Only the equations 4 with $b_3=-1$ and 5 of this list, whose first integrals are autonomous, are known.
Eq. 5 of List \ref{Darboux1} has been found in \cite{h87}, its integrals and a linearizing transformation are presented in \cite{as99}. 

The equation 4 with an arbitrary parameter $b_3$ satisfies all four integrability condi\-tions (\ref{zsoh},\ref{a10}), but it does not have generalized symmetries of the form (\ref{i2},\ref{nondsym}) as well as symmetries similar to eq. \eqref{s2_tz}. It can be proved that eq. 4 possesses $N$-point integrals with $2\le N\le7$ iff $b_3^2=1$. The case $b_3=-1$ is presented in \cite{h79}, the first integrals of this equation in different forms are given in \cite{s10,h11}.

\begin{Spisok}\label{Darboux2}{\bf (Darboux integrable equations: in all cases, one of two $N$-point first integrals is such that $N=2$)}
\begin{enumerate}
\item \label{ur1dar2}
\[u_{n+1,m+1}u_{n,m+1}+b_2(u_{n+1,m}+d)(u_{n,m}+d)=0,\quad b_2\neq0,\]
\[W_1=(-b_2)^{-m}(u_{n+2,m}-u_{n,m})(u_{n+1,m}-u_{n-1,m}),\]
\[W_2=\left(\frac{\sqrt{-b_2}(u_{n,m}+d)}{u_{n,m+1}}\right)^{(-1)^n}\]

\item \label{ur2dar2}
\[u_{n,m}u_{n,m+1}(u_{n+1,m+1}+b_1u_{n+1,m})+u_{n,m}+b_1u_{n,m+1}=0,\quad b_1^2=1,\]
\[W_1=(-b_1)^{m}\frac{u_{n+1,m}u_{n,m}+1}{u_{n,m}},\]
\[W_2=\frac{(u_{n,m+2}+b_1u_{n,m+1})(u_{n,m}+b_1u_{n,m-1})}{(u_{n,m+2}-u_{n,m})(u_{n,m+1}-u_{n,m-1})}\]

\item \label{ur3dar2}
\[u_{n+1,m+1}u_{n,m+1}-u_{n+1,m}u_{n,m}+u_{n,m+1}-u_{n,m}=0,\]
\[W_1=u_{n,m}(u_{n+1,m}+1),\]
\[W_2=\frac{(u_{n,m+2}-u_{n,m})(u_{n,m+1}-u_{n,m-1})}{(u_{n,m+2}-u_{n,m+1})(u_{n,m}-u_{n,m-1})}\]

\item 
\[u_{n+1,m+1}u_{n,m}-u_{n+1,m}u_{n,m+1}+u_{n+1,m}-u_{n,m}=0,\]
\[W_1=\frac{u_{n+1,m}-u_{n,m}}{u_{n,m}-u_{n-1,m}},\qquad W_2=\frac{u_{n,m+1}-1}{u_{n,m}}\]

\item 
\[(u_{n+1,m+1}-u_{n+1,m}+b_4)(u_{n,m+1}-u_{n,m}+b_4)=d^2,\quad d\neq0,\]
 \[W_1=u_{n+1,m}-u_{n-1,m},\]
\[W_2=(-1)^n\frac{u_{n,m+1}-u_{n,m}+b_4-d}{u_{n,m+1}-u_{n,m}+b_4+d}\]

\item 
\[u_{n+1,m+1}(a_2u_{n,m}+u_{n,m+1})+a_2^2u_{n+1,m}u_{n,m}=0,\quad a_2\neq0,\]
 \[W_1=a_2^{-3m}u_{n+1,m}u_{n,m}u_{n-1,m},\]
\[W_2=d^{-n}\frac{a_2 u_{n,m}-du_{n,m+1}}{a_2du_{n,m}-u_{n,m+1}},\ \ d=-\frac{1+\sqrt3 i}2,\ \ i^2=-1\]

\item 
\[u_{n+1,m+1}(a_2u_{n,m}+u_{n,m+1})+u_{n+1,m}(a_2^2u_{n,m}-a_2u_{n,m+1})=0,\quad a_2\neq0,\]
 \[W_1=(-a_2^2)^{-m}u_{n+1,m}u_{n-1,m},\]
\[W_2=i^{n}\frac{a_2 u_{n,m}+iu_{n,m+1}}{a_2iu_{n,m}+u_{n,m+1}},\ \ i^2=-1\]

\item 
\[u_{n+1,m+1}(a_2u_{n,m}+u_{n,m+1})+u_{n+1,m}((a_2^2-d^2)u_{n,m}+a_2u_{n,m+1})=0,\quad d\neq0,\]
 \[W_1=\frac{u_{n+1,m}}{u_{n-1,m}},\qquad
W_2=(-1)^{n}\frac{(a_2+d) u_{n,m}+u_{n,m+1}}{(a_2-d)u_{n,m}+u_{n,m+1}}\]

\item 
\[u_{n+1,m+1}u_{n,m+1}+u_{n+1,m}u_{n,m}+c_4=0,\]
 \[W_1=(-1)^m(2u_{n+1,m}u_{n,m}+c_4),\qquad
W_2=\left(\frac{u_{n,m+1}}{u_{n,m-1}}\right)^{(-1)^n}\]
\end{enumerate}
\end{Spisok}
\bigskip

The equations \ref{ur2dar2} of the last list with $b_1=1$ and $b_1=-1$ are equivalent up to a non-autonomous point transformation to each other and to an equation presented in \cite{as99}. Eq. \ref{ur3dar2} coincides up to an autonomous M\"obius transformation (\ref{mob}) with an example of \cite{s10}. Some of equations of Lists \ref{Darboux1} and \ref{Darboux2} may be related by the non-invertible transformations which, however, are not discussed here, see e.g. examples in \cite{s10,sgr11}.

The first integrals $W_2$ of the form $W_2=\phi^{(-1)^n}$ can be rewritten in the form \eq{\hat W_2=(-1)^n\hat \phi, \qquad \hat \phi=\frac{1+\phi}{1-\phi},\label{W2_step}} see examples given in eq. 3 of List \ref{Darboux1} and in eqs. 1 and 9 of List \ref{Darboux2}. 

The non-autonomous first integrals $W_2$ of the form $$W_2=\left(\sqrt[k]{1}\right)^n\Omega$$ can be rewritten in the autonomous form \seqs{\hat W_2=W_2^k=\Omega^k,} see examples given in eqs. 1,2 and 6 of List \ref{Darboux1} and in eqs. 5,6,7 and 8 of List \ref{Darboux2}. The form of $W_2$ we use is simpler in a sense, and the transition from $\hat W_2$ to $W_2$ is not single-valued. The same can be said about the first integrals $W_1$ given in eqs. 6 of List \ref{Darboux1} and 9 of List \ref{Darboux2}.

The equations of List \ref{Darboux2} are more convenient from the viewpoint of finding their solutions due to the existence of a 2-point first integral, see a detailed discussion in Section \ref{line}. 
This is not an easy problem to construct solutions of Darboux integrable equations using transforms given in eqs. (\ref{darb1},\ref{darb2}). There may exist some more efficient linearizing transformations convenient for the construction of exact solutions. As it has been said above, one example with such a transformation is given by eq. 5 of List \ref{Darboux1}. Another example, i.e. the discrete Burgers equation, is discussed at the end of Section \ref{sineeq}. Examples of this kind together with exact solutions are considered in \cite{a11,as99,s10}. The search for such effective linearizing transformations for the other equations of Lists \ref{Darboux1} and \ref{Darboux2} is an open problem.

\subsection{Equations linearizable via a 2-point first integral}\label{line}

In this section we search for equations of the class \eqref{ovid} which have at least one $2$-point first integral. We show that all such equations are equivalent to a linear non-autonomous discrete equation. As a result we obtain Lists \ref{sp_i1} and \ref{sp_i2}, see below. The equations of List \ref{Darboux2} have two first integrals one of which is $2$-point one, and therefore all equations of this list are contained in Lists \ref{sp_i1} and \ref{sp_i2}. Moreover, Lists \ref{sp_i1} and \ref{sp_i2} will be complete, while List \ref{Darboux2} is just a collection of equations of a certain type.

The class of equations \eqref{ovid} is not symmetric under the transformation $n\leftrightarrow m$. For this reason, we discuss separately two cases corresponding to 2-point first integrals in different directions.

\subsubsection{Case 1}\label{line_s1}

Here we consider 2-point first integrals in the second direction (or in the $m$-direction) which are of the form: \begin{equation}\label{lin_1}(T_1-1)W_2=0,\quad W_2=w_n^{(2)}(u_{n,m},u_{n,m+1}),\end{equation}
where $\partial_{u_{n,m}}W_2\neq0,\partial_{u_{n,m+1}}W_2\neq0$ for all $n.$ An equation \eqref{ovid}, possessing such simple first integral, is {\it equivalent} to the relation \eqref{lin_1} and, therefore, to the equation $W_2=\kappa_m$, where $\kappa_m$ is an arbitrary $m$-dependent function of integration. 

For example, if eq. \eqref{ovid} can be expressed in the form
\eq{\label{t1phi}T_1\phi=\alpha\phi+\beta,\quad \phi=\frac{\nu_1u_{n,m+1}+\nu_2u_{n,m}+\nu_3}{\nu_4u_{n,m+1}+\nu_5u_{n,m}+\nu_6},\quad \alpha\neq0,}
where $\alpha,\beta,\nu_j$ are constant coefficients, then there exists a first integral defined by: 
\eq{W_2=\left\{\begin{array}{l} \phi+n\beta,\quad \alpha=1;\\ \alpha^{-n}\left(\phi+\frac{\beta}{\alpha-1}\right),\quad\alpha\neq1.\end{array}\right.\label{vid_W2}}
This statement is proved by direct calculation. It turns out that all equations \eqref{ovid} possesing a first integral \eqref{lin_1} can be rewritten in the form \eqref{t1phi}, see below.
In case when $W_2$ is defined by \eqref{vid_W2}, the equation $W_2=\kappa_m$ can be rewritten as  \eq{u_{n,m+1}+\mu _{n,m}u_{n,m}+\eta_{n,m}=0\label{lin1_u}.} So, in this case, an equation of the form \eqref{ovid} is {\it equivalent} to a non-autonomous and non-homogeneous linear equation \eqref{lin1_u}.

For any equation of List \ref{sp_i1} below, there is a relation of the form 
\eq{\label{phi_dl}T_1\phi=\frac{\delta_1\phi+\delta_2}{\delta_3\phi+\delta_4},}
shown in the list, and the function $\phi$ has the form shown in eq. \eqref{t1phi}. By using an autonomous linear-fractional transformation of the function $\phi$, we reduce the relation \eqref{phi_dl} to the form  \eqref{t1phi}, and that transformation changes in the formula for $\phi$ the coefficients $\nu_j$ only. So, all equations of List \ref{sp_i1} are equivalent to a linear equation \eqref{lin1_u}. 

Equations of the following list are defined by some relationships for the coefficients $a_j,b_j,c_j$ of eq. \eqref{ovid},
and all the equations have the following restriction: $$a_1=b_1=c_1=0.$$ 

\newpage
\begin{Spisok}{\bf (equations possessing a 2-point first integral of the form \eqref{lin_1})}\\
\label{sp_i1}
\begin{enumerate}
\item \vspace{-0.5cm}
\seqs{a_3\neq0,\quad a_2\neq b_3,\quad a_4\neq c_3,\quad a_2c_3\neq a_3c_2, \\
b_2=\frac{a_3c_2(a_2-b_3)+a_2(b_3a_4-a_2c_3)}{a_3(a_4-c_3)}, \\
b_4=\frac{a_3c_2-c_3a_2+a_4b_3}{a_3}, \\ c_4=\frac{a_3c_2(a_4-c_3)+c_3(a_2c_3-b_3a_4)}{a_3(a_2-b_3)};
 }
$$T_1\phi=\frac{(a_2-b_3)(a_2c_3-a_3c_2)}{(\phi+a_2)(a_4-c_3)}-b_3,\quad\phi=\frac{u_{n,m+1}a_3(a_2-b_3)+c_3a_2-a_4b_3}{u_{n,m}(a_2-b_3)+a_4-c_3} $$
\item 
\seqs{a_3\neq0,\quad b_3\neq 0,\quad a_3c_4\neq a_4c_3, \\
a_2=b_3,\quad b_2=\frac{b_3^2}{a_3},\quad b_4=\frac{a_4b_3}{a_3},\quad c_2=\frac{c_3b_3}{a_3};}
$$T_1\phi=-\frac{c_3\phi+c_4a_3}{\phi+a_4},\quad\phi=u_{n,m}b_3+u_{n,m+1}a_3 $$
\item 
\seqs{a_2\neq0,\quad b_3\neq 0, \\
a_3=0,\quad c_3=\frac{a_4b_3}{a_2},\quad b_4=\frac{a_2(b_3c_2+a_4b_2)-a_4b_2b_3}{a_2^2}, \\ c_4=\frac{a_4(a_2c_2(b_3+a_2)-a_4b_2b_3)}{a_2^3}; }
$$T_1\phi=-\phi\frac{b_3}{a_2}-b_2,\quad \phi=\frac{u_{n,m+1}a_2^2+c_2a_2-a_4b_2}{u_{n,m}a_2+a_4} $$
\item 
\seqs{a_2^2\neq a_3b_2 ,\\
a_4=c_3,\quad b_3=a_2,\quad b_4=c_2,\quad c_4=\frac{2a_2c_2c_3-c_3^2b_2-a_3c_2^2}{a_2^2-b_2a_3} ;}
$$T_1\phi=-\frac{a_2\phi+b_2}{a_3\phi+a_2},\quad \phi=\frac{u_{n,m+1}(a_2^2-a_3b_2)+c_2a_2-b_2c_3}{u_{n,m}(a_2^2-a_3b_2)+c_3a_2-a_3c_2} $$
\item 
\seqs{a_3\neq0,\quad a_2b_3\neq a_3b_2,\\
a_4=c_3,\quad b_4=\frac{c_3b_3}{a_3},\quad c_2=\frac{c_3a_2}{a_3},\quad c_4=\frac{c_3^2}{a_3} ;}
$$T_1\phi=-\frac{b_3\phi+b_2a_3}{\phi+a_2},\quad \phi=\frac{u_{n,m+1}a_3+c_3}{u_{n,m}} $$
\end{enumerate}
\end{Spisok}
\bigskip

For example, any equation of the form 
\[u_{n+1,m+1}(a_2u_{n,m}+a_3u_{n,m+1})+u_{n+1,m}(b_2u_{n,m}+b_3u_{n,m+1})=0,\quad a_2b_3\neq a_3b_2,\] 
is a particular case of eq. 3 (if $a_3=0$) or of eq. 5 (if $a_3\ne 0$) of List \ref{sp_i1}. 
Eqs. 6,7 and 8 of List \ref{Darboux2} are of this form too.

\begin{Theorem}\label{th_i1}
A nondegenerate and nonlinear equation \eqref{ovid} has a first integral \eqref{lin_1} if and only if it belongs to List \ref{sp_i1}. Any equation of List \ref{sp_i1} is equivalent to a linear equation of the form \eqref{lin1_u}. 
\end{Theorem}

 The function $W_2$ satisfies the system of equations \eq{Y_1 W_2=Y_{-1}W_2=Y_2 W_2=Y_{-2}W_2=0. \label{sys_W2}} Analyzing conditions of the solvability of this system, we are led to an algebraic system of equations for the coefficients of eq. \eqref{ovid}. Then we collect for List \ref{sp_i1} all solutions of that algebraic system, such that corresponding equation of the form \eqref{ovid} is nondegenerate and nonlinear. \qed\bigskip

Directly from the system \eqref{sys_W2}, we can find an autonomous function $\phi$ instead of the first integral $W_2$. Necessary relationships for $\phi$ are shown in List \ref{sp_i1} for all equations. Then, using $\phi$, we can introduce an explicit dependence on $n$ and construct a first integral $W_2$ in the way described above. 


\subsubsection{Case 2}

Here a first integral is of the form: 
\begin{equation}\label{lin_2}(T_2-1)W_1=0,\quad W_1=w_m^{(1)}(u_{n,m},u_{n+1,m}),\end{equation}
where $\partial_{u_{n,m}}W_1\neq0,\partial_{u_{n+1,m}}W_1\neq0$ for all $m.$ An equation \eqref{ovid}, possessing such first integral, is {\it equivalent} to the relation \eqref{lin_2} and hence to the equation $W_1=\kappa_n$.

For example, if there are the relations 
\eq{\label{t2phi}T_2\phi=\alpha\phi+\beta,\quad \phi=\frac{\nu_1u_{n+1,m}u_{n,m}+\nu_2u_{n+1,m}+\nu_3u_{n,m}+\nu_4}{\nu_5u_{n,m}+\nu_6},\quad \alpha\neq0,}
with constant coefficients $\alpha,\beta,\nu_j$, then we can construct a first integral as follows: 
\eq{W_1=\left\{\begin{array}{l} \phi+m\beta,\quad \alpha=1;\\ \alpha^{-m}\left(\phi+\frac{\beta}{\alpha-1}\right),\quad\alpha\neq1.\end{array}\right.\label{vid_W1}}
Moreover, the equation $W_1=\kappa_n$ can be rewritten in the form: 
$$\nu_1u_{n+1,m}u_{n,m}+\nu_2u_{n+1,m}+\hat\mu_{n,m}u_{n,m}+\hat\eta_{n,m}=0 .$$
By using a non-autonomous M\"obius transformation (\ref{MOBnm}) of $u_{n,m}$, the last equation can be expressed as:
\eq{u_{n+1,m}+\mu _{n,m}u_{n,m}+\eta_{n,m}=0\label{lin2_u}.} 
So, in this case, an equation of the form \eqref{ovid} is {\it equivalent} to a non-autonomous and non-homogeneous linear equation of the form \eqref{lin2_u}.

For any equation of List \ref{sp_i2} below, we define a function $\phi$ which satisfies relations of the form \eqref{t2phi}. For this reason, there is a first integral defined by \eqref{lin_2} in every case, and all equations of the list are equivalent to a linear equation \eqref{lin2_u}. As well as in the case of List \ref{sp_i1}, the equations of List \ref{sp_i2} are given by some relations for the coefficients $a_j,b_j,c_j$ of eq. \eqref{ovid}. 

\newpage
\begin{Spisok}{\bf (equations possessing a 2-point first integral of the form \eqref{lin_2})}\\
\label{sp_i2}
\begin{enumerate}
\item \vspace{-0.5cm}
\seqs{a_3\neq0,\quad b_1\neq 0,\quad b_2\neq 0,\quad |a_3-a_2|+|b_1c_4-c_2b_2|\neq0, \\
a_1=\frac{b_1a_3}{b_2}, \quad a_4=\frac{a_2b_2}{b_1},\quad b_3=\frac{a_2b_2}{a_3},\quad b_4=\frac{a_2b_2^2}{b_1a_3},\\
c_1=\frac{b_1(b_2^2c_2+a_3b_2c_2-a_3b_1c_4)}{b_2^3}, \quad
c_3=\frac{b_1b_2c_4+a_3b_2c_2-a_3b_1c_4}{b_2^2}; }
\seqs{T_2\phi=-\phi\frac{b_2}{a_3}-b_2c_2-a_3c_2+\frac{b_1a_3c_4}{b_2},\\ \phi=\frac{u_{n+1,m}u_{n,m}b_1b_2a_3+u_{n+1,m}a_2b_2^2+a_3(b_1c_4-b_2c_2)}{u_{n,m}b_1+b_2} }
\item
\seqs{a_3\neq0,\quad b_2\neq 0, \\
a_1=0,\quad a_2=0,\quad a_4=\frac{b_4a_3}{b_2}, \quad b_1=0,\quad b_3=0,\quad c_1=0,\quad c_3=\frac{c_2a_3}{b_2}; }
$$T_2\phi=-\frac{b_2}{a_3}(\phi+c_4),\quad\phi=u_{n+1,m}u_{n,m}b_2+u_{n+1,m}b_4+u_{n,m}c_2 $$
\item
\seqs{a_4\neq0,\quad b_3\neq 0,\quad b_4\neq 0, \\
a_1=0,\quad a_2=\frac{a_4b_3}{b_4},\quad a_3=0, \quad b_1=0,\quad b_2=0,\\
c_2=\frac{c_1b_4^2+b_3a_4c_3-a_4b_4c_1}{b_3b_4} ,\quad
c_4=\frac{b_3b_4c_3+a_4b_3c_3-a_4b_4c_1}{b_3^2} ;}
$$T_2\phi=-\frac{b_4}{a_4}(\phi+c_1),\quad\phi=\frac{u_{n+1,m}b_3^2+b_3c_3-c_1b_4}{u_{n,m}b_3+b4} $$
\item
\seqs{a_1\neq0,\quad b_1\neq 0,\quad |b_3|+|c_3|\neq0, \\
 	a_2=\frac{a_1b_3}{b_1},\quad a_3=0, \quad a_4=0, \quad b_2=0,\quad b_4=0,\quad c_2=\frac{a_1c_3}{b_1},\quad c_4=0;}
$$T_2\phi=-\frac{b_1}{a_1}(\phi+c_1),\quad\phi=\frac{u_{n+1,m}u_{n,m}b_1+u_{n+1,m}b_3+c_3}{u_{n,m}} $$
\item
\seqs{a_2\neq0,\quad b_3\neq 0, \\
a_1=0,\quad a_3=0, \quad a_4=0, \quad b_1=0,\quad b_2=0,\quad b_4=0,\quad c_2=\frac{a_2c_3}{b_3},\quad c_4=0;}
$$T_2\phi=-\frac{b_3}{a_2}(\phi+c_1),\quad\phi=\frac{u_{n+1,m}b_3+c_3}{u_{n,m}} $$
\end{enumerate}
\end{Spisok}
\bigskip

For example, eqs. 2,3 and 9 of List \ref{Darboux2} are the particular cases of equations presented in List \ref{sp_i2}. 

\begin{Theorem}
A nondegenerate and nonlinear equation of the form \eqref{ovid} has a first integral defined by \eqref{lin_2} if and only if it belongs to List \ref{sp_i2}. Any equation of List \ref{sp_i2} is equivalent to a linear equation of the form  \eqref{lin2_u} up to a non-autonomous M\"obius transformation (\ref{MOBnm}) of its solutions $u_{n,m}$.\label{th_i2}
\end{Theorem}

A proof of this theorem is completely analogous to the proof of Theorem \ref{th_i1}. 

\paragraph{\small Remark.} It seems to be true that for Darboux integrable autonomous equations \eqref{gF}, any $N$-point non-autonomous first integral can be rewritten as an $N+1$-point autonomous one, sf. for example \cite{h08} for the semidiscrete case. In this paper, due to the remark \eqref{W2_step}, all non-autonomous integrals have the form \eqref{vid_W2} or \eqref{vid_W1}. The following formulae: 
\eq{\widetilde W_2=\left\{\begin{array}{l}T_2(W_2)-W_2,\ \alpha=1;\\ T_2(W_2)/W_2, \ \alpha\neq1,\end{array}\right.\quad \widetilde W_1=\left\{\begin{array}{l}T_1(W_1)-W_1,\ \alpha=1;\\ T_1(W_1)/W_1, \ \alpha\neq1,\end{array}\right.} provide the autonomous first integrals corresponding to \eqref{vid_W2} and \eqref{vid_W1}.

The use of the non-autonomous first integrals is, however, well motivated. In Section \ref{darboux} we construct the first integrals of the lowest possible order, simplifying in this way the calculation. In Section \ref{line} we use the simplest $2$-point first integrals. Consideration of the non-autonomous first integrals essentially enlarges the set of possible examples. As it can be seen from Lists \ref{sp_i1} and \ref{sp_i2}, all ten examples, except for their particular cases, have the non-autonomous first integral.

\subsection{Additional results}\label{addit}
 
Here we specify the nature of integrability of some known equations. 

First of them is the $Q_V$ equation introduced by Viallet in \cite{v09} which sometimes is called the Adler equation \cite{a98} with free coefficients. All equations of the Adler-Bobenko-Suris list are contained in the $Q_V$ equation. 

This equation is defined by the following conditions for the polylinear function $F$ of eq. \eqref{gF}:
\seqs{
F (u_{n,m}, u_{n+1,m}, u_{n,m+1}, u_{n+1,m+1}) = F (u_{n+1,m}, u_{n,m}, u_{n+1,m+1}, u_{n,m+1})
\\= F (u_{n,m+1}, u_{n+1,m+1}, u_{n,m}, u_{n+1,m}) ,} and it depends on $7$ arbitrary constant parameters. As it has been shown in \cite{x09}, the $Q_V$ equation has the generalized symmetries \eqref{two_sym} for all values of these parameters. We are interested in the intersection of our class \eqref{ovid} and the $Q_V$ equation, which has the form 
\eqs{
  (u_{n,m} u_{n+1,m} &+ u_{n,m+1} u_{n+1,m+1}) k_1 + 
    (u_{n,m} u_{n+1,m+1} + u_{n+1,m} u_{n,m+1}) k_2   \label{viollet}\\
 & +(u_{n,m} + u_{n+1,m} + u_{n,m+1} + u_{n+1,m+1}) k_{3} + k_{4} = 0. }
If $k_1=k_2=0$, then this equation is linear.

\begin{Theorem} Any equation of the form \eqref{viollet}, such that $k_1\neq0$ or $k_2\neq0$, is Darboux integrable or degenerate or it is equivalent to a linear equation up to an autonomous point transformation $\tilde u_{n,m} = \varphi (u_{n,m})$. 
\end{Theorem}

\paragraph{Proof.} Considering all possible cases, it is easy to check that eq. \eqref{viollet} is either degenerate or equivalent up to a linear transformation $\hat u_{n,m}=\alpha u_{n,m}+\beta$ to one of the following equations: one of eqs. \eqref{l1} or one of eqs. 1,2,4 of List \ref{Darboux1} or eq. 9 of List \ref{Darboux2}. \qed\bigskip

The next two equations do not belong to the class \eqref{ovid}, but a result will be interesting and general.

The $Q_V$ equation is generalized by a class of polylinear equations defined in \cite{x09} by:
\eqs{\label{u4}
F (u_{n,m}, u_{n+1,m}, u_{n,m+1}, u_{n+1,m+1}) &= \pi_1 F(u_{n+1,m}, u_{n,m}, u_{n+1,m+1}, u_{n,m+1})
\\&= \pi_2 F (u_{n,m+1}, u_{n+1,m+1}, u_{n,m}, u_{n+1,m}) ,
}
where $\pi_1 = \pm 1$, $\pi_2 = \pm 1$. All equations of the class \eqref{u4} also have the generalized symmetries \eqref{two_sym}, see \cite{x09}. In addition to the $Q_V$ equation, we have here (up to the transformation $n\leftrightarrow m$) two more equations: 
\eqs{  (u_{n+1,m}& u_{n,m+1} u_{n+1,m+1} + u_{n,m} u_{n,m+1} u_{n+1,m+1} -
     u_{n,m} u_{n+1,m} u_{n+1,m+1} \\&- u_{n,m} u_{n+1,m} u_{n,m+1}) k_1 + 	
  (u_{n,m} u_{n+1,m} - u_{n,m+1} u_{n+1,m+1}) k_2 \\&+
    (u_{n,m} + u_{n+1,m} - u_{n,m+1} - u_{n+1,m+1}) k_{3} = 0 , \label{u5}	
}
\eqs{
  (u_{n+1,m}& u_{n,m+1} u_{n+1,m+1} - u_{n,m} u_{n,m+1} u_{n+1,m+1} -
     u_{n,m} u_{n+1,m} u_{n+1,m+1}\\& + u_{n,m} u_{n+1,m} u_{n,m+1}) k_1 + 
  (u_{n,m} u_{n+1,m+1} - u_{n+1,m} u_{n,m+1}) k_2 \\&+
    (u_{n,m} - u_{n+1,m} - u_{n,m+1} + u_{n+1,m+1}) k_{3} = 0 . \label{u6}	
}
It has been shown in \cite{ly11} that both equations (\ref{u5}) and (\ref{u6}) are equivalent to a particular case of the $Q_V$ equation. We have here a more interesting result:

\begin{Theorem} 
Any equation of the form (\ref{u5}) or (\ref{u6}) is equivalent to a linear equation up to an autonomous point transformation $\tilde u_{n,m} = \varphi (u_{n,m})$.
\end{Theorem}

\paragraph{Proof.} We do not consider the trivial case $k_1=k_2=k_3=0$. Both classes \eqref{u5} and \eqref{u6} are invariant under the autonomous M\"obious transformations (\ref{mob}). By using such transformations, we can always make $k_1=0$ and $k_2k_3=0$ in both equations (\ref{u5}) and (\ref{u6}). In this way we get either a linear equation or one of eqs. \eqref{l1}. \qed\bigskip

As it is has been said in \cite{ly11}, the following particular case of the $Q_V$ equation: 
\eqs{
  &(u_{n+1,m} u_{n,m+1} u_{n+1,m+1} + u_{n,m} u_{n,m+1} u_{n+1,m+1} +
     u_{n,m} u_{n+1,m} u_{n+1,m+1} \\&+ u_{n,m} u_{n+1,m} u_{n,m+1}) k_1 + 
  (u_{n,m} + u_{n+1,m} + u_{n,m+1} + u_{n+1,m+1}) k_{2} = 0  \label{u7}	
} is transformed into an equation of the form \eqref{u5} by the transformation $\hat u_{n,m}=(-1)^m u_{n,m}$. This means that eq. \eqref{u7} is equivalent to a linear equation up to a non-autonomous point transformation $\tilde u_{n,m} = \varphi_{m} (u_{n,m})$.

\section{Conclusion}

We have in this paper the following results, presented in Section \ref{sec3}:
\begin{enumerate}
\item We have performed the exhaustive classification of sine-Gordon type integrable equations of the class \eqref{ovid}, more precisely, of equations possessing properties of List \ref{prop}, Section \ref{sec21}. The resulting equations are collected in List \ref{sineql}, Section \ref{sineeq}. In this way we have shown that there are no more new sine-Gordon type equations in the class \eqref{ovid}.
\item We have found eq. \eqref{tzit} with a nonstandard symmetry structure which pretends to be a new example of integrable discrete equation. Its most simple generalized symmetries (\ref{s1_tz},\ref{s2_tz}) have a more complicated form than \eqref{two_sym}, unlike all equations of List \ref{sineql}. Both of them are not autonomous, moreover, the symmetries (\ref{s2_tz},\ref{s3_tz}) have arbitrary $n$-dependent two-periodic coefficients. The generalized symmetry \eqref{s2_tz} in itself is possibly new example of an integrable differential-difference equation of the Itoh-Narita-Bogoyavlensky type. Eq. \eqref{tzit} should be not lineari\-zable, at least not Darboux integrable, as we have checked that it does not have any $N$-point first integral with $2\le N\le 7$.
\item When carrying out the generalized symmetry classification, we find some of Darboux integrable equations in a natural way. We have collected such equations in Lists \ref{Darboux1} and \ref{Darboux2}, Section \ref{darboux}. Most of those equations are possibly new, and most of equations have the non-autonomous first integrals which are considered in the literature, probably, for the first time. 
\item In Section \ref{line} we consider equations which can be rewritten as a non-autonomous linear equation by using a $2$-point first integral. We enumerate all equations of this kind within the class (\ref{ovid}) and put them in Lists \ref{sp_i1} and \ref{sp_i2}. 
\item In Section \ref{addit} we specify the nature of integrability of some known equations introduced in \cite{v09,x09}, in particular, of the intersection of $Q_V$ and the class \eqref{ovid}. We show that all nondegenerate equations of this intersection are either Darboux integrable or linearizable via an autonomous point transformation.
\end{enumerate}

The class \eqref{ovid} is a particular case of the general class of polylinear equations of the form (\ref{gF}), which depends on $16$ arbitrary constants. For this reason, the work in all directions, we considered in this paper, can be continued.

\paragraph{Acknowledgments.}
This work  has been supported by the Russian Foundation for Basic Research (grant
numbers: 10-01-00186-a, 10-01-00088-a, 12-01-92602-KO-a, 11-01-97005-r-povolzhie-a). We thank I.T. Habibullin for a useful discussion.


\begin{thebibliography}{99}

\bibitem{a98}
V.E. Adler, \emph{B\"acklund transformation for the Krichever-Novikov equation,} Inter\-nat. Math. Res. Notices (1998), no. 1, 1--4.

\bibitem{a11}
V.E. Adler, \emph{On a discrete analog of the Tzitzeica equation,} {\tt arXiv:1103.5139}.

\bibitem{abs03}
V.E. Adler,  A.I. Bobenko and Yu.B. Suris,
\emph{Classification of integrable equations on quad-graphs. The consistency approach},
Commun. Math. Phys. {\bf 233}  (2003) 513--543.

\bibitem{asy00}
V.E. Adler, A.B. Shabat and R.I. Yamilov, \emph{Symmetry approach to the integrability problem}, Teoret. Mat. Fiz. {\bf 125}:3 (2000) 355--424 (in Russian); English transl. in Theor. Math. Phys. {\bf 125}:3 (2000) 1603--1661.

\bibitem{as99} V.E. Adler and S.Ya. Startsev, \emph{Discrete analogues of the Liouville equation}, Teoret. Mat. Fiz. {\bf 121}:2 (1999) 271–284 (in Russian); English transl. in Theor. Math. Phys. {\bf 121}:2 (1999) 1484--1495.

\bibitem{bo88}
O.I. Bogoyavlensky, \emph{Integrable discretizations of the KdV equation}, Phys. Lett. A {\bf 134}:1 (1988) 34--38. 

\bibitem{ggh11}R.N. Garifullin, E.V. Gudkova and I.T. Habibullin, \emph{Method for searching higher symmetries for quad-graph equations}, J. Phys. A: Math. Theor. {\bf 44} (2011) 325202. 

\bibitem{sgr11}B. Grammaticos, A. Ramani, C. Scimiterna and R. Willox, \emph{Miura transformations and the various guises of integrable lattice equations}, J. Phys. A: Math. Theor. {\bf 44} (2011) 152004. 

\bibitem{h05}I.T. Habibullin, \emph{Characteristic algebras of fully discrete hyperbolic type equations}, SIGMA  Symmetry Integrability Geom. Methods Appl. {\bf 1} (2005) 023.

\bibitem{h08} I. Habibullin, N. Zheltukhina and A. Pekcan, \emph{On the classification of Darboux integrable chains}, J. Math. Phys. {\bf 49} (2008) 102702 (39 pages).

\bibitem{h11}I. Habibullin, N. Zheltukhina and A. Sakieva, \emph {Discretization of hyperbolic type Darboux integrable equations preserving integrability}, J. Math. Phys. {\bf 52} (2011) 093507 (12 pages).

\bibitem{h04} J. Hietarinta, \emph{A new two-dimensional lattice model that is ‘consistent around a cube’}, J. Phys. A: Math. Gen. {\bf 37} (2004) L67--L73.

\bibitem{hv07}
J. Hietarinta and C. Viallet, \emph{Searching for integrable lattice maps using factorization},
J. Phys. A: Math. Theor. {\bf 40} (2007) 12629--12643.

\bibitem{h79} R. Hirota, \emph{Nonlinear partial difference equations. V. Nonlinear equations reducible to linear equations}, J. Phys. Soc. Japan {\bf 46} (1979) 312--319.

\bibitem{h87} R. Hirota, \emph{Discrete two-dimensional Toda molecule equation}, J. Phys. Soc. Japan {\bf 56} (1987) 4285--4288.

\bibitem{i75} Y. Itoh, \emph{An $H$-theorem for a system of competing species}, Proc. Japan Acad. {\bf 51} (1975) 374--379.

\bibitem{lpsy08}
D. Levi, M. Petrera, C. Scimiterna and R. Yamilov, \emph{On Miura transformations and Volterra-type equations associated with the Adler-Bobenko-Suris equations}, SIGMA Symmetry Integrability Geom. Methods Appl. {\bf 4} (2008) 077, 14 pages.

\bibitem{ly97} D. Levi and R. Yamilov, \emph{Conditions for the existence of higher
symmetries of evolutionary equations on the lattice}, J. Math. Phys. {\bf 38}
(1997) 6648--6674.

\bibitem{ly09}
D. Levi and R.I. Yamilov, \emph{The generalized symmetry method for discrete
equations}, J. Phys. A: Math. Theor. {\bf 42} (2009) 454012 (18pp).

\bibitem{ly11}D. Levi and R.I. Yamilov, \emph{Generalized symmetry integrability test for
discrete equations on the square lattice}, J. Phys. A: Math. Theor. {\bf 44} (2011) 145207 (22pp).

\bibitem{MS11}
A. G. Meshkov and V. V. Sokolov, \emph{Hyperbolic equations with third-order symmetries}, Teoret. Mat. Fiz. {\bf 166}:1 (2011) 51–67  (in Russian); English transl. in Theor. Math. Phys. {\bf 166}:1 (2011) 43--57.

\bibitem{msy87}
A.V. Mikhailov, A.B. Shabat and R.I. Yamilov, \emph{The symmetry
approach to the classification of nonlinear equations. Complete lists of
integrable systems}, Uspekhi Mat. Nauk {\bf 42}:4 (1887) 3--53 (in Russian);
English transl. in Russian Math. Surveys {\bf 42}:4 (1987) 1--63.

\bibitem{mwx11}
A. Mikhailov, J.P. Wang and P. Xenitidis, \emph{Recursion operators, conservation laws, and integrability conditions for difference equations}, Teoret. Mat. Fiz. {\bf 167}:1 (2011) 23--49 (in Russian); English transl. in Theor. Math. Phys.  {\bf 167}:1 (2011) 421--443.

\bibitem{na82}
K. Narita, \emph{Soliton solution to extended Volterra equation}, J. Phys. Soc. Japan {\bf 51}:5 (1982) 1682--1685.

\bibitem{rj06} A. Ramani, N. Joshi, B. Grammaticos and T. Tamizhmani, \emph{Deconstructing an integrable lattice equation}, J. Phys. A: Math. Gen. {\bf 39} (2006) L145--L149.

\bibitem{rh07} O.G. Rasin and P.E. Hydon, \emph{Conservation laws for integrable difference equations}, J. Phys. A: Math. Theor. {\bf 40} (2007) 12763--12773.

\bibitem{s10} S.Ya. Startsev, \emph{On non-point invertible transformations of difference and 
diffe\-rential-difference equations}, SIGMA Symmetry Integrability Geom. Methods Appl. {\bf 6} (2010) 092, 14 pages. 

\bibitem{s11} A.K. Svinin, \emph{On some integrable lattice related by the Miura-type transformation to the Itoh–Narita–Bogoyavlenskii lattice}, J. Phys. A: Math. Theor. {\bf 44} (2011) 465210.

\bibitem{t907} G. Tzitzeica, \emph{Sur une nouvelle classe de surfaces,} Rendiconti del Circolo Mate\-matico di Palermo {\bf 25}:1 (1907) 180--187.

\bibitem{v09}
C.M. Viallet, \emph{Integrable lattice maps: $Q_V$, a rational version of $Q_4$}, Glasgow Math. J. {\bf 51} (2009) 157--163.

\bibitem{x09}
P. Xenitidis, \emph{Integrability and symmetries of difference equations: the Adler-Bobenko-Suris case},
In Proc. of 4th Workshop ``Group Analysis of Differential Equations and Integrable Systems'', 2009,  226--242,
{\tt arXiv:0902.3954}.

\bibitem{xp09}
P.D. Xenitidis and V.G. Papageorgiou, \emph{Symmetries and integrability of discrete equations
defined on a black--white lattice}, J. Phys. A: Math. Theor. {\bf 42} (2009) 454025 (13pp).

\bibitem{y83}
R.I. Yamilov, \emph{Classification of discrete evolution equations},
Uspekhi Mat. Nauk {\bf 38}:6 (1983) 155--156 (in  Russian).

\bibitem{y06}
R. Yamilov, \emph{Symmetries as integrability criteria for
differential difference equa\-tions}, J. Phys. A: Math. Gen. {\bf 39} (2006)
R541--R623.

\bibitem{zs79} A.V. Zhiber and A.B. Shabat, \emph{The Klein-Gordon equation with nontrivial group,} Dokl. Akad. Nauk SSSR {\bf 247}:5 (1979) 1103--1107 (in Russian).

\bibitem{ZS01}
A.V. Zhiber and V.V. Sokolov, \emph{Exactly integrable hyperbolic equations of Liouville type}, Uspekhi Mat. Nauk {\bf 56}:1 (2001) 63–106  (in Russian);
English transl. in Russian Math. Surveys {\bf 56}:1 (2001) 61--101.

\end{thebibliography}
\end{document}